# Theoretical Analysis of the Coherence Properties of a Grating

Ivan A. Vartanyants[1,*]

[1]*Deutsches Elektronen-Synchrotron DESY, Notkestr. 85, 22607 Hamburg, Germany*

**September 6, 2024**

## Abstract

In this work we aim to clarify theoretically the spatial coherence properties of the x-ray beam in the focal plane after interaction with a variable line space (VLS) grating. Assuming that the VLS grating is oriented horizontally, we are interested in the coherence properties of the beam in the vertical dispersion direction. We first consider a fully spatially coherent x-ray beam illuminating the grating. We show that the spatial coherence properties in the focal plane depend on the bandwidth of the incoming radiation. Being fully spatially coherent up to the VLS grating, the spatial coherence properties degrade in the focal plane of the VLS grating. We attribute this to coupling of the spatial and frequency components in the focal plane of the grating. Next, we examine partially coherent x-ray beams incident on a VLS grating. We assume that this radiation is of the Gaussian Schell-model type and obtain an analytical expression for the spatial coherence properties of the beam in the focal plane of the VLS grating for such a field. Next, we consider a monochromator setting that is provided by installing slits in the focal plane of the VLS grating and examine the degree of coherence in this case. Finally, we evaluate the degree of coherence at different apertures of the exit slits, assuming coherent illumination of the grating and a bandwidth of $1.7 \cdot 10^{-4}$.

[*]Corresponding author: Ivan.Vartaniants@desy.de

# I. Introduction

The existing and planned fourth generation synchrotron sources based on multi-bend achromat lattice design provides an extreme rise in the spatial coherence properties of the x-ray beams [1-5]. These sources are highly coherent for tender and high energies and are diffraction limited in the soft x-ray range [6]. This means that the x-ray beams will be fully spatially coherent starting from the source and it also means that special care has to be taken to design a beamline in the soft x-ray energy range that will preserve high spatial coherence properties of the x-ray beam. We will assume that the soft x-ray beam is generated by an undulator and that the spatial coherence of such a beam is quite high and is about 80%-90% [6]. However, it is well known that the spectral coherence from such an undulator is quite low and spectrally we have a rather incoherent beam.

It is well known, that the quality of optics may influence the parameters of the beam incident on the sample positioned in the experimental hutch. For example, surface defects may alter the wavefront profile of the beam [7-9] and vibration of optics can degrade the coherence properties of the beam [10,11]. In the following, we leave all these questions aside and consider a stable beamline with ideal optics. We plan to investigate how the spatial coherence properties of the incident soft x-ray beam may be affected by scattering on a grating.

Gratings and variable line space (VLS) gratings have been widely used since the 1980s as an important part of the soft x-ray monochromator [12-18]. In addition to the grating, the monochromator typically also accommodates slits that choose a certain bandwidth of radiation and focusing mirrors. In principle, VLS gratings are themselves focusing devises, but in the case the focal length of the monochromator with the VLS grating is not sufficient, sometimes it is necessary to add a focusing mirror to such a device.

Gratings and, in particular, VLS gratings, were quite widely discussed in the past [13-18], however, the spatial coherence properties of these gratings were left almost unexplored [19]. If we assume that the VLS grating is oriented horizontally then we assume that in the horizontal direction the spatial coherence properties will be preserved after propagation through the grating. At the same time, the coherence properties in the vertical, dispersion direction, are not so obvious. In the recent paper [20] it was stated, by performing numerical simulations, that the spatial coherence properties of the x-ray beam with a small incoming bandwidth are degraded in the dispersion direction when scattering from grating. Here, we performed a theoretical analysis of the scattering from a grating and analyzed this degradation of the spatial coherence. As we will see in the following, this is also related to the bandwidth of radiation incoming on the grating. In the monochromatic case, the spatial coherence is unaffected in the vertical plane and we will get perfect coherence.

We first analyze the spatial coherence properties of the x-ray beam generated by a coherent source with a small frequency bandwidth. This situation will be especially important for the 4[th] generation synchrotron sources. As a fully spatially coherent beam, such an x-ray beam will preserve its properties up to the grating as each frequency propagates independently. However, in the focal plane of the grating, we observed degradation of the spatial coherence. We next assumed that the x-ray beam incident on the grating is partially coherent and has a special form of a Gaussian Schell-model beam. We analyze theoretically the propagation of such a beam through the VLS grating and deduce the spatial coherence properties of the beam in the focal plane of the grating. Subsequently, we include the exit slit aperture of the VLS grating monochromator and analyze the spatial coherence as a function of slit opening.

We start in the next section with the basic coherence theory.

## II. Basics of coherence theory

The first order coherence is defined through the mutual coherence function (MCF) [21,22]

$$\Gamma(\boldsymbol{r_1}, \boldsymbol{r_2}, t_1, t_2) = \langle E^*(\boldsymbol{r_1}, t_1) E(\boldsymbol{r_2}, t_2) \rangle. \qquad (1)$$

It describes correlations between two values of the electric field, $E(\boldsymbol{r_1}, t_1)$ and $E(\boldsymbol{r_2}, t_2)$, at different points $\boldsymbol{r_1}$ and $\boldsymbol{r_2}$ and times $t_1$ and $t_2$. The brackets $\langle \ldots \rangle$ denote the ensemble average; for stationary sources the ensemble average coincides with the time average. For stationary radiation, Eq. (1) may be written in the following form

$$\Gamma(\boldsymbol{r_1}, \boldsymbol{r_2}, \tau) = \langle E^*(\boldsymbol{r_1}, t_1) E(\boldsymbol{r_2}, t_1 + \tau) \rangle_T, \qquad (2)$$

where $t_2 = t_1 + \tau$ and averaging is performed over time. A convenient measure of spatial coherence is the normalized MCF [22]

$$\gamma(\boldsymbol{r_1}, \boldsymbol{r_2}, \tau) = \frac{\Gamma(\boldsymbol{r_1}, \boldsymbol{r_2}, \tau)}{[\langle I(\boldsymbol{r_1}, t) \rangle]^{1/2} [\langle I(\boldsymbol{r_2}, t) \rangle]^{1/2}} \qquad (3)$$

which is called the complex degree of coherence (CDC) and is in general a complex function. The modulus of the CDC is ranging from zero to one and depends on the pinhole separation $\Delta \boldsymbol{r} = \boldsymbol{r_2} - \boldsymbol{r_1}$ and are determined in the classical Young's experiment.

We next introduce the cross-spectral density (CSD) function of a stationary source in spatial-frequency domain, which is defined as

$$W(\boldsymbol{r_1}, \boldsymbol{r_2}, \omega) = \langle E^*(\boldsymbol{r_1}, \omega) E(\boldsymbol{r_2}, \omega') \rangle \delta(\omega - \omega'), \qquad (4)$$

where $\omega$ is the frequency of radiation and $\delta(\omega - \omega')$ is the Dirac delta function. Equation (4), in fact, states, that for stationary radiation each frequency is propagating independently. It is well established that the MCF and CSD for stationary radiation are connected by a Fourier transform [21,22]

$$\Gamma(r_1, r_2, \tau) = \frac{1}{2\pi} \int W(r_1, r_2, \omega) e^{-i\omega\tau} d\omega . \tag{5}$$

The spectral density of the radiation field is obtained from Eq. (4) when two points $r_1$ and $r_2$ coincide, $r = r_1 = r_2$

$$S(\mathbf{r}, \omega) = W(r, r, \omega) = \langle |E(r, \omega)|^2 \rangle . \tag{6}$$

As we will see further, we will be interested in interference effects under conditions close to geometrical symmetry, so for us $\tau$ will be close to zero. In this case, we can approximate our correlation functions by the following expressions [22]

$$\begin{aligned}\Gamma(r_1, r_2, \tau) &\approx J(r_1, r_2) exp(-i\omega_0\tau) , \\ \gamma(r_1, r_2, \tau) &\approx j(r_1, r_2) exp(-i\omega_0\tau) ,\end{aligned} \tag{7}$$

where $\omega_0$ is the carrier frequency. We can use an approximation of Eqs. (7) if the following condition is satisfied

$$|\tau| \ll 1/\Delta\omega , \tag{8}$$

where $\Delta\omega$ is the bandwidth of the radiation. We will see in the following that condition (8) is very well satisfied in our case. Following [21,22], we will call the correlation function $J(r_1, r_2)$ as a mutual intensity function (MIF) at points $r_1$ and $r_2$ and $j(r_1, r_2)$ as a complex degree of coherence (CDC) similar to $\gamma(r_1, r_2, \tau)$. In fact, the quantities $J(r_1, r_2)$ and $j(r_1, r_2)$ are the equal time correlation functions [22]

$$\begin{aligned}J(r_1, r_2) &\equiv \Gamma(r_1, r_2, 0) = \langle E^*(r_1, t)E(r_2, t)\rangle_T , \\ j(r_1, r_2) &\equiv \gamma(r_1, r_2, 0) = \frac{J(r_1, r_2)}{[J(r_1, r_1)]^{1/2}[J(r_2, r_2)]^{1/2}} .\end{aligned} \tag{9}$$

In the case when condition (8) is satisfied, we have from Eqs. (5) and (9)

$$J(r_1, r_2) = \frac{1}{2\pi} \int W(r_1, r_2, \omega) d\omega . \tag{10}$$

From Eqs. (9) we can also obtain the intensity of radiation field $I(\mathbf{r})$ when two points $r_1$ and $r_2$ coincide, $r = r_1 = r_2$

$$I(\mathbf{r}) \equiv \Gamma(r, r, 0) = \langle E^*(r, t)E(r, t)\rangle_T . \tag{11}$$

Another convenient measure of coherence is the degree of coherence $\varsigma^{DC}$, which characterizes the coherence properties of the wavefield by a single number and can be introduced as [23-25]

$$\varsigma^{DC} = \frac{\int |J(\mathbf{r}_1, \mathbf{r}_2)|^2 \, d\mathbf{r}_1 d\mathbf{r}_2}{[\int I(\mathbf{r}) d\mathbf{r}]^2}. \qquad (12)$$

The values of the parameter $\varsigma^{DC}$ lie in the range $0 \leq \varsigma^{DC} \leq 1$, where $\varsigma^{DC} = 1$ and $\varsigma^{DC} = 0$ correspond to fully coherent and incoherent radiation, respectively.

It is well known [21], that the wavefield is completely coherent when the MIF can be represented as a product of two complex valued amplitudes

$$J(\mathbf{r}_1, \mathbf{r}_2) = \langle E^*(\mathbf{r}_1, t) E(\mathbf{r}_2, t) \rangle = |E(\mathbf{r}_1)||E(\mathbf{r}_2)| exp[i(\alpha_2 - \alpha_1)], \qquad (13)$$

where $\alpha_1$ and $\alpha_2$ are the phases of the complex valued amplitudes $E(\mathbf{r}_1)$ and $E(\mathbf{r}_2)$. Direct substitution of expression (13) in Eqs. (9) and (12) shows that the modulus of CDC $|j(\mathbf{r}_1, \mathbf{r}_2)|$ and the degree of coherence $\varsigma^{DC}$ are equal to unity, that proves that such a field is fully coherent.

The Gaussian Schell-model (GSM) is an often-used model [24,25] that represents the radiation for a real x-ray beam based on the following approximations: the source is modelled as a plane two-dimensional source, the source is spatially uniform, i.e. the CDC depends only on the difference $\Delta \mathbf{r}$, and the CDC $j(\Delta \mathbf{r}, \omega)$ as well as intensity $I(\mathbf{r})$ are Gaussian functions. In the frame of the GSM MIF, intensity, and CDC are defined as (see, for example, [22])

$$J(\mathbf{r}_1, \mathbf{r}_2) = \sqrt{I(\mathbf{r}_1)} \sqrt{I(\mathbf{r}_2)} j(\Delta \mathbf{r}), \qquad (14)$$

$$I(\mathbf{r}) = I_0 exp\left(-\frac{x^2}{2\sigma_x^2} - \frac{y^2}{2\sigma_y^2}\right), \qquad (15)$$

$$j(\Delta \mathbf{r}) = exp\left[-\frac{(x_2 - x_1)^2}{2\xi_x^2} - \frac{(y_2 - y_1)^2}{2\xi_y^2}\right], \qquad (16)$$

where $I_0$ is a normalization constant, $\sigma_{x,y}$ is the r.m.s. source size and $\xi_{x,y}$ is the transverse coherence length in the source plane in the x- and y-directions, respectively. One of the important features of this model is that the MIF is separable into two transverse directions,

$$J(\mathbf{r}_1, \mathbf{r}_2, \omega) = J(x_1, x_2) J(y_1, y_2) \qquad (17)$$

The same is valid for the degree of coherence as defined by Eq. (12)

$$\varsigma^{DC} = \varsigma_x^{DC} \varsigma_y^{DC}, \qquad (18)$$

where in each transverse direction $i = x, y$ we have [24,25]

$$\varsigma_i^{DC} = \frac{\xi_i/\sigma_i}{\sqrt{4 + (\xi_i/\sigma_i)^2}}. \tag{19}$$

Under the paraxial approximation, a wavefield at one plane with distribution $E_{z_0}(\mathbf{r}', \omega)$ can be propagated to the wavefield at another plane at a distance z, which has a distribution $E_z(\mathbf{r}, \omega)$, and can be obtained as

$$E_z(\mathbf{r}, \omega) = \int E_{z_0}(\mathbf{r}', \omega) P_z(\mathbf{r} - \mathbf{r}') d\mathbf{r}', \tag{20}$$

where $P_z(\mathbf{r} - \mathbf{r}')$ is a Green function or propagator that is given by

$$P_z(\mathbf{r} - \mathbf{r}') = \frac{k}{i(2\pi)z} \exp\left[\frac{ik}{2z}(\mathbf{r} - \mathbf{r}')^2\right], \tag{21}$$

where $k = 2\pi/\lambda$ is the absolute value of the wavevector and $\lambda$ is the wavelength of x-ray radiation.

### III. Scattering on a grating

We will start with the scattering on a grating in reflection geometry (see Fig. 1) following the outline of the known approach [26] (see Appendix for more details). We also assume that the grating size $D_N = d_0 N$, where $d_0$ is the spacing of the grooves of the grating and $N$ is the number of grooves, is much larger than the illuminating field and we will consider scattering to the *n*-th order. Taking all this into account, we may obtain the following relation (see Eq. (AI, 14) in Appendix) for the amplitude of the field in the focal plane of the VLS grating

$$E_{gr}(q_x, \omega) = E_0 r_{gr}(h_n) \int E_{in}(x', \omega) e^{-i(q_x - h_n) \cdot x'} dx'. \tag{22}$$

Here, $q_x$ is the momentum transfer that is defined in Eq. (AI, 13), $r_{gr}(h_n)$ is the Fourier component of the reflection function of one period, $h_n = (2\pi/d_0)n$, $E_0$ is the amplitude of the field and $E_{in}(x', \omega)$ is the amplitude of the incoming beam that is, in principle, fluctuating. The integration in Eq. (22) is performed along the grating.

Now, we would like to take a closer look at the term

$$q_x - h_n = \left(\frac{\omega}{c}\right)(\sin \beta_n + \sin \alpha) - h_n. \tag{23}$$

This equation is specific for gratings. Importantly, it says that if we consider a monochromatic Gaussian incoming field centered on the carrier frequency $\omega_0$ it will be scattered to angles $\beta_n$ defined by the following condition

$$k_0(\sin \beta_n + \sin \alpha) = h_n, \qquad (24)$$

where $k_0 = \omega_0/c$. Equation (24) is known as a grating equation. We will consider now that the frequency ω is slightly shifted from the carrier frequency ω₀ by Δω and we have a corresponding shift of the angle $\beta_n$ by Δβ

$$\begin{aligned}\omega &= \omega_0 + \Delta\omega, \Delta\omega \ll \omega_0 ; \\ \beta &= \beta_n + \Delta\beta, \Delta\beta \ll \beta_1 ; \\ \Delta\omega &= \omega - \omega_0, \Delta\beta = \beta - \beta_n.\end{aligned} \qquad (25)$$

Substituting all of this into Eq. (23) we obtain

$$q_x - h_n = \left(\frac{\omega_0 + \Delta\omega}{c}\right)[\sin(\beta_n + \Delta\beta) + \sin \alpha] - h_n.$$

Now, using the known relation

$$\sin(\beta_n + \Delta\beta) = \sin \beta_n \cos \Delta\beta + \cos \beta_n \sin \Delta\beta \approx \sin \beta_n + \Delta\beta \cos \beta_n ,$$

the condition that only small terms up to the first order may contribute to the final result, and taking into consideration the grating Eq. (24), we obtain

$$q_x - h_n = k_0 \Delta\beta \cos \beta_n + h_n(\Delta\omega/\omega_0). \qquad (26)$$

This is a very important result as it says that if the frequency of the incoming beam $\omega$ is shifted from the carrier frequency $\omega_0$ by $\Delta\omega$, then as a response of the spectrometer there will be an angular shift $\Delta\beta$ of the maximum intensity for a certain order $n$.

Now, taking into account that $f$ is the focal distance of the VLS grating at which the exit slits are positioned, we can introduce a shift in space $\Delta x = x - x_n$ ($x_n$ is the position where the central frequency $\omega_0$ will give its main contribution at the $n$-th order) that is given by an angle $\Delta\beta = \Delta x/f$. Substituting all this into Eq. (22) we, finally, obtain for the amplitude of the wavefield scattered to the exit slit position

$$E_{gr}(\Delta x, \Delta\omega) = E_0 r_{gr}(h_n) \int E_{in}(x', \omega) e^{-i[q_x + q_\omega]x'} dx', \qquad (27)$$

where

$$\begin{aligned}q_x &= k_0 \cos \beta_n \left(\Delta x/f\right), \\ q_\omega &= h_n(\Delta\omega/\omega_0).\end{aligned} \qquad (28)$$

This is an important equation that provides spatial-frequency coupling after scattering from a grating. We note here that for a monochromatic wave with frequency $\omega = \omega_0$, when $\Delta\omega = 0$ or $\omega = \omega_0$ we have

$$E_{gr}(\Delta x, \omega_0) = E_0 r_{gr}(h_n) \int E_{in}(x', \omega_0) e^{-iq_x x'} dx' . \tag{29}$$

This means that each peak of frequency $\omega = \omega_0$ after scattering from the grating will have a certain shape in the $x$-domain.

## IV. Spatial coherence properties in the focal plane of the VLS grating. Spectrometer configuration

We will evaluate now the MIF and other correlation functions, including the degree of coherence for two cases when the incoming field on the grating is completely coherent and when the incoming field is partially coherent.

We will, first, express the CSD function $W_{gr}(x_1, x_2, \omega)$ (see Eq. (4)) in the focal plane in the $n$-th order of the VLS grating by putting the field values $E_{gr}(x, \omega)$ for the grating according to Eq. (27) into the expression (4)

$$\begin{aligned} W_{gr}(q_{x_1}, q_{x_2}, \omega) &= \langle E^*_{gr}(x_1, \omega) E_{gr}(x_2, \omega) \rangle = \\ &= W_0 \iint \langle E^*_{in}(x', \omega) E_{in}(x'', \omega) \rangle e^{-i(q_{x_2} x'' - q_{x_1} x')} e^{-iq_\omega(x'' - x')} dx' dx'' , \end{aligned} \tag{30}$$

where $W_0$ is the term in which we will further incorporate all nonessential pre-integral factors and

$$q_{x_{1,2}} = k_0 \cos\beta_n \left[ (x_{1,2} - x_n)/f \right]. \tag{31}$$

As we already mentioned before, the incoming field on the grating $E_{in}(x, \omega)$ is fluctuating, which is why we kept correlations inside the integral. Assuming now that the incoming radiation is cross-spectral pure [21,22], we obtain for the CSD function of the incoming field

$$\begin{aligned} W_{in}(x', x'', \omega) &= \langle E^*_{in}(x', \omega) E_{in}(x'', \omega) \rangle = \langle E^*_{in}(x') E_{in}(x'') \rangle \langle E^*_{in}(\omega) E_{in}(\omega) \rangle \\ &= J_{in}(x', x'') \langle |E_{in}(\omega)|^2 \rangle = J_{in}(x', x'') S_{in}(\omega) , \end{aligned} \tag{32}$$

where we used definitions of Eqs. (6) and (9).

We will consider now that the spectrum of the incoming beam is Gaussian and has the following form

$$S_{in}(\omega) = S_0 exp\left[-\frac{(\Delta\omega)^2}{2(\sigma_{in}^{\omega})^2}\right] = S_0 exp\left[-\frac{1}{2(\sigma_{in}^{\omega'})^2}\left(\frac{\Delta\omega}{\omega_0}\right)^2\right], \quad (33)$$

where $S_0$ is a normalization constant, $\Delta\omega = \omega - \omega_0$ and

$$\sigma_{in}^{\omega'} = \frac{\sigma_{in}^{\omega}}{\omega_0}. \quad (34)$$

Now, assuming that the condition in Eq. (8) is well satisfied in our case and substituting our results of Eqs. (30, 32) in expression (10) for the MIF we obtain

$$J_{gr}(q_{x_1}, q_{x_2}) = \frac{1}{2\pi}\int W_{gr}(x_1, x_2, \omega)\, d\omega =$$
$$= J_0 \iint dx' dx'' J_{in}(x', x'') e^{-i(q_{x_2}x'' - q_{x_1}x')} * \quad (35)$$
$$* \int d\omega\, S_{in}(\omega) e^{-iq_\omega(x''-x')},$$

where $J_0$ is the term in which we will further incorporate all nonessential pre-integral factors.

Now, by integrating over $\omega$ in Eq. (35) and using an integral given in Eq. (AII, 1) we obtain

$$\int_{-\infty}^{\infty} d\omega S_{in}(\omega) e^{-iq_\omega(x''-x')} = \sqrt{2\pi}\sigma_{in}^{\omega'} S_0 exp\left\{-\frac{(x''-x')^2}{2l_c^2}\right\}, \quad (36)$$

where

$$l_c = \frac{1}{h_n \sigma_{in}^{\omega'}} = \frac{1}{h_n}\left(\frac{\omega_0}{\sigma_{in}^{\omega}}\right). \quad (37)$$

As we see from these results, due to integration over $\omega$ we now obtain a "partially coherent" beam with the spatial coherence length $l_c$. In the case of a monochromatic beam $\sigma_\omega \Rightarrow 0$, we have for a coherence length $l_c \Rightarrow \infty$ and we have a fully coherent situation. In all other cases $l_c$ is finite. We should also emphasize that this is valid for a situation with completely open slits in the so-called spectrometer configuration of the grating.

Now, substituting the result of the integration (36) into Eq. (35) we obtain for the MIF

$$J_{gr}(q_{x_1}, q_{x_2}) = J_0 \iint dx' dx'' J_{in}(x', x'')\, exp\left\{-\frac{(x''-x')^2}{2l_c^2}\right\} e^{-i(q_{x_2}x'' - q_{x_1}x')}. \quad (38)$$

Now we will perform analysis for two distinct cases: illumination of a grating by a *completely coherent* and *partially coherent* incoming field.

### a. Illumination of a grating by a spatially coherent field

We will first assume that the incoming field is *completely coherent*. In this case, in our geometry we have from Eqs. (14-16), that the spatial coherence length of the incoming beam $\xi_{in} \to \infty$ and we have for $J_{in}(x', x'')$

$$J_{in}(x_1, x_2) = \sqrt{I_{in}(x_1)}\sqrt{I_{in}(x_2)}, I_{in}(x) = I_0 exp\left(-\frac{x^2}{2(\sigma_{in}^x)^2}\right). \quad (39)$$

We also have to consider, that in our geometry (see Fig. 1) the footprint of the beam along the grating has its root mean square (rms) value $\sigma_{in}^{x\,\prime} = \sigma_{in}^x/\cos\alpha$ and we would have to substitute $\sigma_{in}^x$ by $\sigma_{in}^{x\,\prime}$ in Eq. (39).

Now substituting Eq. (39) in Eq. (38) we obtain

$$\begin{aligned}J_{gr}(q_{x_1}, q_{x_2}) &= \\ &= J_0 \iint dx' dx'' \sqrt{I_{in}(x')}\sqrt{I_{in}(x'')} exp\left\{-\frac{(x''-x')^2}{2l_c^2}\right\} e^{-i(q_{x_2}x''-q_{x_1}x')} = \\ &= J_0 \iint dx' dx'' e^{-(ax'^2+ax''^2-2bx'x'')} e^{-i(q_{x_2}x''-q_{x_1}x')} \\ &= J_0 e^{-(\alpha q_{x_1}^2 + \alpha q_{x_2}^2 - 2\beta q_{x_1} q_{x_2})},\end{aligned} \quad (40)$$

where

$$a = \frac{1}{2}\left(\frac{1}{2(\sigma_{in}^{x\,\prime})^2} + \frac{1}{l_c^2}\right); b = \frac{1}{2l_c^2} \quad (41)$$

and parameters $\alpha$ and $\beta$ are defined in Eq. (AII, 3). Following Eq. (AII, 4) we can write for the MIF for coherent illumination of the VLS grating at the focal plane position

$$J_{gr}(\Delta x_1, \Delta x_2) = J_0 exp\{-[\tilde{\alpha}\Delta x_1^2 + \tilde{\alpha}\Delta x_2^2 - 2\tilde{\beta}\Delta x_1 \Delta x_2]\}, \quad (42)$$

where parameters $\tilde{\alpha}$ and $\tilde{\beta}$ are defined as

$$\tilde{\alpha} = \frac{(2+q_c^2)}{\sigma_1^2(4+q_c^2)}; \tilde{\beta} = \frac{2}{\sigma_1^2(4+q_c^2)} \quad (43)$$

and

$$q_c = \frac{l_c}{\sigma_{in}^{x\,\prime}}; \sigma_1 = \frac{1}{k_0}\left(\frac{f}{\sigma_{in}^x}\right)\left(\frac{\cos\alpha}{\cos\beta_n}\right). \quad (44)$$

Following now the results of Eqs. (AII, 6 - AII, 10), we obtain for the intensity $I_{gr}(\Delta x)$ of the field in the focal plane of the VLS grating (see Eq. (11))

$$I_{gr}(\Delta x) = I_0 exp\left[-\frac{(\Delta x)^2}{2(\sigma_I^c)^2}\right], \tag{45}$$

where

$$\sigma_I^c = \frac{1}{2\sqrt{\tilde{\alpha} - \tilde{\beta}}} = \frac{\sigma_1\sqrt{4 + q_c^2}}{2q_c}. \tag{46}$$

We further calculate the CDC $j_{gr}(x_1, x_2)$ that is defined in Eqs. (9) and we have according to Eq. (AII, 8)

$$j_{gr}(x_1, x_2) = exp\left[-\frac{(x_2 - x_1)^2}{2L_c^2}\right], \tag{47}$$

where we have introduced the coherence length $L_c$ in the focus of the grating as

$$L_c = \frac{1}{\sqrt{2\tilde{\beta}}} = \frac{1}{2}\sigma_1\sqrt{4 + q_c^2}. \tag{48}$$

For small values of $q_c \ll 1$, we have for the coherence length $L_c$

$$L_c \cong \sigma_1$$

and for large values, $q_c \gg 1$, we have

$$L_c \cong {(q_c\sigma_1)}/{2}.$$

Finally, for the spatial degree of transverse coherence (see Eqs. (12 and AII, 10)), we obtain

$$\zeta_{gr}^{DC} = \left[\frac{\tilde{\alpha} - \tilde{\beta}}{\tilde{\alpha} + \tilde{\beta}}\right]^{1/2} = \frac{q_c}{\sqrt{4 + q_c^2}}. \tag{49}$$

Here, for small values of $q_c \ll 1$, we have

$$\zeta_{gr}^{DC} \cong {(q_c}/{2)}\left[1 - \frac{1}{2}\left(\frac{q_c}{2}\right)^2\right]$$

and for large values, $q_c \gg 1$, we have

$$\zeta_{gr}^{DC} \cong 1 - \frac{1}{2}\left(\frac{2}{q_c}\right)^2$$

or as $q_c \Rightarrow \infty$ the degree of coherence $\zeta_{gr}^{DC} \Rightarrow 1$ and we would obtain a completely spatially coherent field. We will see now at which conditions we could obtain such a field. Taking now into account Eqs. (37, 44) we have for the parameter $q_c$ the following expression

$$q_c = \frac{l_c}{\sigma_{in}^{x'}} = \frac{1}{h_n \sigma_{in}^{\omega'} \sigma_{in}^{x'}} = \frac{1}{2\pi}\left(\frac{d_0}{n\sigma_{in}^x}\right)\left(\frac{\omega_0}{\sigma_{in}^\omega}\right)\cos\alpha. \tag{50}$$

We immediately see from this equation that the parameter $q_c \Rightarrow \infty$ only for a monochromatic wave with $\sigma_{in}^\omega \Rightarrow 0$. In all other cases the parameter $q_c$ will have a finite value.

It is interesting to note, that the expression for the degree of coherence in Eq. (49) is similar to the Gaussian Schell-model expression in Eq. (19) with a substitution $q_c = \xi_i/\sigma_i$. In the case of coherent illumination of a grating the parameter $l_c$ plays the same role as the transverse coherence length in each direction $\xi_i$.

Here, we have obtained an interesting result. Initially, we assumed that a completely spatially coherent beam is incoming on the grating, however, after scattering on a grating, we obtained a partially coherent beam in the direction of dispersion of the grating. In the transverse direction the coherence properties of the beam will be conserved. Now we will examine the case when the grating is illuminated by *a spatially partially coherent* beam.

### b. *Illumination of a grating by a spatially partially coherent field*

We will now consider that the incoming field is *spatially partially coherent* and is of Gaussian Schell-model type. In this case, in our geometry, we have from Eqs. (14-16) for $J_{in}(x', x'')$

$$J_{in}(x_1, x_2) = \sqrt{I_{in}(x_1)}\sqrt{I_{in}(x_2)}\, j_{in}(x_2 - x_1),$$
$$I_{in}(x) = I_0 \exp\left[-\frac{x^2}{2(\sigma_{in}^x)^2}\right], \quad (51)$$
$$j_{in}(\Delta x) = \exp\left[-\frac{(x_2 - x_1)^2}{2\xi_{in}^2}\right].$$

We also have to take into account, that in our geometry (see Fig. 1) the rms size of the beam along the grating is $\sigma_{in}^{x\,\prime} = \sigma_{in}^x/\cos\alpha$ and the same is valid for the coherence length of the incoming beam $\xi_{in}' = \xi_{in}/\cos\alpha$.

Substituting now Eqs. (51) into Eq. (38) we obtain in this case for the MIF in the focal plane of the VLS grating

$$J_{\text{gr}}(q_{x_1}, q_{x_2}) =$$

$$= J_0 \iint \sqrt{I_{in}(x')I_{in}(x'')} j_{in}(x''$$

$$- x') exp\left[-\frac{(x''-x')^2}{2l_c^2}\right] e^{-i(q_{x_2}x''-q_{x_1}x')} dx'dx'' = \quad (52)$$

$$= J_0 \iint dx'dx'' e^{-(ax'^2+ax''^2-2bx'x'')} e^{-i(q_{x_2}x''-q_{x_1}x')}$$

$$= J_0 e^{-(\alpha q_{x_1}^2 + \alpha q_{x_2}^2 - 2\beta q_{x_1} q_{x_2})},$$

where the parameters $q_{x_{1,2}}$ and $l_c$ are defined in Eqs. (31, 37). In Eq. (52) the parameters $a$ and $b$ have the value

$$a = \frac{1}{2}\left[\frac{1}{2(\sigma_{in}^{x'})^2} + \frac{1}{(\xi'_{in})^2} + \frac{1}{l_c^2}\right]; b = \frac{1}{2}\left[\frac{1}{(\xi'_{in})^2} + \frac{1}{l_c^2}\right]. \quad (53)$$

and parameters $\alpha$ and $\beta$ are defined in Eq. (AII, 3). Following now Eq. (AII, 4) we can write for the MIF for the partial coherent illumination of the VLS grating at the focal plane position

$$J_{\text{gr}}(\Delta x_1, \Delta x_2) = J_0 exp\{-[\tilde{\alpha}\Delta x_1^2 + \tilde{\alpha}\Delta x_2^2 - 2\tilde{\beta}\Delta x_1 \Delta x_2]\}, \quad (54)$$

where parameters $\tilde{\alpha}$ and $\tilde{\beta}$ are defined as

$$\tilde{\alpha} = \frac{2q_t^2 + 2q_c^2 + q_t^2 q_c^2}{\sigma_1^2(4q_t^2 + 4q_c^2 + q_t^2 q_c^2)}; \tilde{\beta} = \frac{2(q_t^2 + q_c^2)}{\sigma_1^2(4q_t^2 + 4q_c^2 + q_t^2 q_c^2)}, \quad (55)$$

and the dimensionless parameter $q_t$ is

$$q_t = \frac{\xi'_{in}}{\sigma_{in}^{x'}} = \frac{\xi_{in}}{\sigma_{in}^{x}}. \quad (56)$$

Following now the results of Eqs. (AII, 6 - AII, 10), we obtain for the intensity $I_{gr}(\Delta x)$ of the field in the focal plane of the VLS grating (see Eq. (11)) in the case of the *partially coherent* illumination

$$I_{gr}(\Delta x) = I_0 exp\left[-\frac{(\Delta x)^2}{2(\sigma_I^{pc})^2}\right], \quad (57)$$

where now for the *partially coherent* case

$$\sigma_I^{pc} = \frac{1}{2\sqrt{\tilde{\alpha} - \tilde{\beta}}} = \frac{\sigma_1}{2q_t q_c}[4q_t^2 + 4q_c^2 + q_t^2 q_c^2]^{1/2}. \quad (58)$$

We further calculate the CDC $j_{gr}(x_1, x_2)$ that is defined in Eqs. (9) (see also Eq. (AII, 8))

$$j_{gr}(x_2 - x_1) = exp\left[-\frac{(x_2 - x_1)^2}{2(L_c^{pc})^2}\right], \tag{59}$$

where now the coherence length $L_c^{pc}$ in the focus of the grating in the case of the *partially coherent* incoming radiation is defined as

$$L_c^{pc} = \frac{1}{\sqrt{2\tilde{\beta}}} = \frac{1}{2}\sigma_1 \left[\frac{4q_t^2 + 4q_c^2 + q_t^2 q_c^2}{q_t^2 + q_c^2}\right]^{1/2}. \tag{60}$$

Now, according to definition of the degree of spatial coherence (see Eq. (12)) and following Eq. (AII, 10) we obtain

$$\zeta_{gr}^{DC} = \left[\frac{\tilde{\alpha} - \tilde{\beta}}{\tilde{\alpha} + \tilde{\beta}}\right]^{1/2} = \frac{q_t q_c}{[4q_t^2 + 4q_c^2 + q_t^2 q_c^2]^{1/2}}. \tag{61}$$

We note here that all expressions from the MIF in Eq. (54) to the degree of coherence in Eq. (61) coincide with our previous results in Eqs. (42 - 49) with the *completely coherent* illumination case when $q_t \Rightarrow \infty$.

## V. Resolution of the spectrometer

Now we can determine the resolution of our spectrometer at the exit slit position. Assuming that we are at the fundamental frequency $\omega = \omega_0$ (see Eq. (29)) and that the incoming beam on the grating is defined by a *spatially coherent* field Eq. (39), we obtain for the scattered intensity at the exit slit position

$$I_{gr}(\Delta x, \omega_0) = |E_0|^2 |r_{gr}(h_n)|^2 \left|\int E_{in}(x', \omega_0) exp(-iq_x x') dx'\right|^2 =$$

$$= |E_0|^2 |r_{gr}(h_n)|^2 exp\left[-\frac{\Delta x^2}{2(\sigma_c^x)^2}\right] \tag{62}$$

with the rms value $\sigma_c^x$

$$\sigma_c^x = \frac{\lambda_0 f}{4\pi \sigma_{in}^x}\left(\frac{\cos\alpha}{\cos\beta_n}\right) = \sigma_1/2. \tag{63}$$

In the case of *partially coherent* illumination the scattered intensity at the exit slit position will be determined as

$$I_{gr}(q_x, \omega_0) = |E_0|^2 |r_{gr}(h_n)|^2 \iint dx' dx'' J_{in}(x', x'') e^{-iq_x(x''-x')}, \tag{64}$$

where $J_{in}(x', x'')$ is defined by Eq. (51). By performing calculations, we, finally, obtain the following result

$$I_{gr}(q_x, \omega_0) = |E_0|^2 |r_{gr}(h_n)|^2 exp\left[-\frac{\Delta x^2}{2(\sigma_{pc}^x)^2}\right], \quad (65)$$

where

$$\sigma_{pc}^x = \frac{\lambda_0 f}{4\pi \sigma_{in}^x}\left(\frac{\cos \alpha}{\cos \beta_n}\right)\sqrt{1 + 1/q_t^2} = \frac{\sigma_1}{2}\frac{\sqrt{1+q_t^2}}{q_t}. \quad (66)$$

We note here, that when $q_t \Rightarrow \infty$ we obtain the previous *completely coherent* case.

To determine the transition from the spatial domain to frequency or energy domains we can use the dispersion relation in Eq. (24)

$$\sin \beta_n + \sin \alpha = \frac{\lambda_0}{d_0} n \quad (67)$$

Taking the derivative of both parts, we obtain

$$\cos \beta_n \, \delta \beta_n = \frac{\lambda_0}{d_0}\left(\frac{\delta \lambda}{\lambda_0}\right) n. \quad (68)$$

Considering that $\delta \beta_n = (x - x_n)/f = \delta x_n/f$ and $\delta \lambda/\lambda_0 = -\delta \omega/\omega_0 = -\delta E/E_0$ we obtain

$$\delta x_n = -\frac{\lambda_0 f}{d_0 \cos \beta_n}\left(\frac{\delta \omega}{\omega_0}\right) n. \quad (69)$$

This equation provides a transition from the spatial domain to frequency domain. The same relation can be obtained directly from Eq. (26) by assuming that $q_x - h_n \equiv 0$.

To determine now the resolution, we should be based on the Rayleigh's principle [26], when the maximum for the wavelength $\lambda_0$ is compared with the minimum of the shifted wavelength $\lambda_0 \pm \delta \lambda$. However, we cannot directly apply Rayleigh's principle as we assume all functions to be Gaussian functions. According to that, we will assume that two lines are resolved if a frequency change by $\delta \omega/\omega_0$ brings a shift of the Gaussian peak by the rms value $\sigma_c^x$ (see Eq. (63)) or $\sigma_{pc}^x$ (see Eq. (66)): $|\delta x_n| = \sigma_c^x$ or $|\delta x_n| = \sigma_{pc}^x$. In fact, this condition is as arbitrary as Rayleigh's principle, however, we will check that it will not bring us to contradiction with the known Rayleigh's principle. As a result, we are coming to the following relation (for the *completely coherent* case)

$$\frac{\delta\omega}{\omega_0} = \frac{\sigma_{res}^\omega}{\omega_0} = \frac{\sigma_1}{2}\frac{d_0 \cos\beta_n}{\lambda_0 f |n|} = \frac{d_0 \cos\alpha}{4\pi |n| \sigma_{in}^x}. \tag{70}$$

Taking into account that the Full Width at Half Maximum (FWHM) of an area illuminated by the incoming beam on the grating is equal to $FWHM = 2\sqrt{2ln2}\,\sigma_{in}^x/\cos\alpha = 2.355\,\sigma_{in}^x/\cos\alpha$, and that the $FWHM = d_0 N$, where $N$ is a number of illuminated grooves on the grating, we obtain from Eq. (70)

$$\frac{\delta\omega}{\omega_0} = \frac{\sigma_{res}^\omega}{\omega_0} = \frac{2.355 d_0}{4\pi |n| FWHM} \sim \frac{1}{|n|N} \tag{71}$$

as it should be for a grating (see for e.g. [26]).

In the *partially coherent* illumination case, Eq. (70) will take the form

$$\frac{\delta\omega}{\omega_0} = \frac{\sigma_{res}^\omega}{\omega_0} = \frac{\sigma_1}{2}\frac{\sqrt{1+q_t^2}}{q_t}\frac{d_0 \cos\beta_n}{\lambda_0 f |n|} = \frac{d_0 \cos\alpha}{4\pi |n| \sigma_{in}^x}\frac{\sqrt{1+q_t^2}}{q_t}.$$

**VI.     Coherence properties in the focal plane of the VLS grating. Monochromator configuration**

Before, we were considering the case of completely open slits in the exit plane of the grating, the so-called spectrometer configuration. Now, we will consider the so-called monochromator configuration, when the exit plane of the grating is equipped with slits of varying sizes (see Fig. 2).

In this case, the amplitude of the field after the exit slits $E_{sl}(\Delta x, \Delta\omega)$ may be obtained by multiplying the amplitude of the field scattered from the grating to a fixed order $E_{gr}(\Delta x, \Delta\omega)$ (see Eq. (27)) by the transmission function of the slits, $T_{sl}(\Delta x)$

$$E_{sl}(\Delta x, \Delta\omega) = T_{sl}(\Delta x) E_{gr}(\Delta x, \Delta\omega), \tag{72}$$

where

$$T_{sl}(\Delta x) = exp\left[-\frac{(\Delta x)^2}{2(\sigma_{sl}^x)^2}\right] \tag{73}$$

and $\Delta x = x - x_n$. In this equation the transmission function of the slits $T_{sl}(\Delta x)$ is defined as a finite window in the spatial domain of the spatial size $\sigma_{sl}^x$. We also assume that the slits are aligned at the center of the corresponding order, by that we can consider that $x_n = 0$ and by that eliminate the sign $\Delta$. We will also consider that measurements of coherence are performed on a detector located at some distance $L$ from the exit slits of the grating (see Fig. 2), which we define as the detector plane with the coordinate $x^D$ and the field value at this position as

$$E_D(x^D, \omega) = \int E_{sl}(x', \omega) P_L(x^D - x')\, dx', \qquad (74)$$

where the one-dimensional (1D) propagator $P_L(x^D - x')$ is defined according to Eq. (21) as

$$P_L(x^D - x') = \sqrt{\frac{k}{i(2\pi)L}} \exp\left[\frac{ik}{2L}(x^D - x')^2\right]. \qquad (75)$$

We will position slits in the focal plane of the VLS grating. In a general case, we may obtain for the MIF at the detector position

$$J_D(x_1^D, x_2^D) = J_0 \iint dx_1 dx_2 T^*(x_1) T(x_2) J_{gr}(x_1, x_2) P_L^*(x_1^D - x_1) P_L(x_2^D - x_2). \qquad (76)$$

Here $J_{gr}(x_1, x_2)$ is the MIF of the field in the focus of the VLS grating in front of the slits (see Eq. (38)). This MIF has a form of Eq. (42) for the *completely coherent* illumination and a form of Eq. (54) in the case of the *partially coherent* illumination of the grating. First, we will write the propagator in Eq. (76) in the following form and assume far-field conditions ($L \gg kd^2$, where d is a typical size)

$$P_L(x^D - x') = \sqrt{\frac{k}{i(2\pi)L}} \exp\left[\frac{ik}{2L}(x^D)^2 - i q_{x^D} x'\right], \qquad (77)$$

where

$$q_{x^D} = \frac{k}{L} x^D \qquad (78)$$

Now by substituting Eqs. (73, 77) in Eq. (76) we obtain for the MIF

$$J_D(x_1^D, x_2^D) =$$
$$= J_0 \exp\left\{i\left(\frac{k}{2L}\right)[(x_2^D)^2 - (x_1^D)^2]\right\} \qquad (79)$$
$$\iint dx_1 dx_2\, \exp\{-[\bar{\alpha} x_1^2 + \bar{\alpha} x_2^2 - 2\bar{\beta} x_1 x_2]\} \exp\left[-i\left(q_{x_2^D} x_2 - q_{x_1^D} x_1\right)\right],$$

where

$$\bar{\alpha} = \tilde{\alpha} + \frac{1}{2(\sigma_{sl}^x)^2} \; ; \; \bar{\beta} = \tilde{\beta} \qquad (80)$$

and the parameters $\tilde{\alpha}$ and $\tilde{\beta}$ where introduced for the *completely coherent* illumination case in Eq. (43) and for the *partially coherent* illumination of the grating in Eq. (55). The integral in Eq. (79) is a type of integral in Eq. (AII, 2) and we can write immediately

$$J_D(x_1^D, x_2^D) = J_0 exp\left\{i\left(\frac{k}{2L}\right)[(x_2^D)^2 - (x_1^D)^2]\right\}$$
$$exp\left[-\left(\bar{\bar{\alpha}} q_{x_1^D}^2 + \bar{\bar{\alpha}} q_{x_2^D}^2 - 2\bar{\bar{\beta}} q_{x_1^D} q_{x_2^D}\right)\right], \qquad (81)$$

where

$$\bar{\bar{\alpha}} = \frac{\bar{\alpha}}{4(\bar{\alpha}^2 - \bar{\beta}^2)} \; ; \; \bar{\bar{\beta}} = \frac{\bar{\beta}}{4(\bar{\alpha}^2 - \bar{\beta}^2)}. \qquad (82)$$

Finally, we can write for the MIF at the detector position in the far-field the following expression

$$J_D(x_1^D, x_2^D) = J_0 exp\left\{i\left(\frac{k}{2L}\right)[(x_2^D)^2 - (x_1^D)^2]\right\}$$
$$exp[-(\hat{\alpha}(x_1^D)^2 + \hat{\alpha}(x_2^D)^2 - 2\hat{\beta} x_1^D x_2^D)], \qquad (83)$$

where

$$\hat{\alpha} = \alpha_0^2 \bar{\bar{\alpha}} \; ; \; \hat{\beta} = \beta_0^2 \bar{\bar{\beta}}; \; \alpha_0^2 = \beta_0^2 = \left(\frac{k}{L}\right)^2. \qquad (84)$$

Now, following the results of Appendix II (see Eqs. (AII, 6-AII, 10)) we have for the intensity of the field on the detector

$$I^D(x) = I_0 exp\left[-\frac{x^2}{2(\sigma_I^D)^2}\right], \qquad (85)$$

where

$$\sigma_I^D = \frac{1}{2\sqrt{\hat{\alpha} - \hat{\beta}}} = \frac{L}{k_0 \sigma_1 q_{sl}}[1 + q_{sl}^2]^{1/2} \qquad (86)$$

and we have introduced a dimensionless parameter $q_{sl}$ as

$$q_{sl} = \sqrt{2}\frac{\sigma_{sl}^x}{\sigma_1}. \qquad (87)$$

It is interesting to note here, that the same result stated in Eq. (86) for the rms value of the intensity of the field at the detector position is valid for the *completely coherent* illumination as well as for the *partially coherent* illumination of the grating. We note here that when the slits are open, $q_{sl} \Rightarrow \infty$, and we have for the intensity on the detector in Eq. (88) the value

$$\sigma_I^D = \frac{L}{k_0 \sigma_1}$$

and when the slits are rather closed, $q_{sl} \Rightarrow 0$, we obtain for the same quantity

$$\sigma_I^D = \frac{L}{k_0 \sigma_1 q_{sl}}.$$

For the CDC $j^D(x_1, x_2)$ on the detector in the case of *completely coherent* illumination of the grating, we obtain according to its definition in Eq. (9) and Eq. (AII, 4)

$$j^D(x_1, x_2) = exp\left\{i\left(\frac{k}{2L}\right)[(x_2^D)^2 - (x_1^D)^2]\right\} exp\left[-\frac{(x_2 - x_1)^2}{2(L_c^D)^2}\right], \tag{88}$$

where

$$L_c^D = \frac{1}{\sqrt{2\hat{\beta}}} = \frac{L}{k_0 \sigma_1 q_{sl}^2} \{[1 + q_{sl}^2][q_c^2 q_{sl}^2 + (4 + q_c^2)]\}^{1/2}. \tag{89}$$

So, in this case the modulus of CDC $|j(\Delta x_1, \Delta x_2)|$ will be given by a Gaussian function with its rms value given by Eq. (89). We note here that for the *completely coherent* illumination of the grating when the slits are open, $q_{sl} \Rightarrow \infty$, we have for $L_c^D$ on the detector in Eq. (92) the value

$$L_c^D = \frac{L q_c}{k_0 \sigma_1}$$

and when the slits are rather closed $\sigma_{sl}^x \Rightarrow 0$ we obtain for the same quantity

$$L_c^D = \frac{L}{k_0 \sigma_1 q_{sl}^2}\sqrt{4 + q_c^2}.$$

For the CDC $j(\Delta x_1, \Delta x_2)$ in the case of *partially coherent* illumination of the grating, we obtain according to its definition in Eq. (9) and Eq. (AII, 4)

$$j_{gr}(\Delta x_1, \Delta x_2) = exp\left\{i\left(\frac{k}{2L}\right)[(x_2^D)^2 - (x_1^D)^2]\right\} exp\left[-\frac{(x_2 - x_1)^2}{2(L_c^D)^2}\right], \tag{90}$$

where

$$L_c^D = \frac{1}{\sqrt{2\hat{\beta}}} = \frac{L}{k_0 \sigma_1 q_{sl}^2}\left\{\frac{[1 + q_{sl}^2][q_t^2 q_c^2 q_{sl}^2 + (4q_t^2 + 4q_c^2 + q_t^2 q_c^2)]}{q_t^2 + q_c^2}\right\}^{1/2}. \tag{91}$$

So, in this case the modulus of CDC $|j(\Delta x_1, \Delta x_2)|$ will be given by a Gaussian function with its rms value given by Eq. (91). We note here that for the *partially coherent* illumination case when the slits are open, $q_{sl} \Rightarrow \infty$, we have for $L_c^D$ on the detector in Eq. (91) the value

$$L_c^D = \frac{L q_t q_c}{k_0 \sigma_1 \sqrt{q_t^2 + q_c^2}}$$

and when the slits are rather closed, $q_{sl} \Rightarrow 0$, we obtain for the same quantity

$$L_c^D = \frac{L}{k_0 \sigma_1 q_{sl}^2} \left[ \frac{4q_t^2 + 4q_c^2 + q_t^2 q_c^2}{q_t^2 + q_c^2} \right]^{1/2}.$$

We note here, that this result coincides with the *completely coherent* illumination case given above when the parameter $q_t \Rightarrow \infty$.

Finally, evaluating the spatial degree of the transverse coherence on the detector position $\varsigma_D^{DC}$ that was defined in Eq. (12) in the case of *completely coherent* illumination of the grating, we obtain

$$\varsigma_D^{DC} = \left[ \frac{\hat{\alpha} - \hat{\beta}}{\hat{\alpha} + \hat{\beta}} \right]^{1/2} = \left[ \frac{4 + q_c^2 + q_c^2 q_{sl}^2}{(4 + q_c^2)(1 + q_{sl}^2)} \right]^{1/2}. \qquad (92)$$

We see from this expression that when the slits are open, $q_{sl} \Rightarrow \infty$, we have for the degree of coherence our previous coherent result stated in Eq. (49) and when the slits are rather closed, $q_{sl} \Rightarrow 0$, we obtain for the degree of coherence just the value of one or hundred percent of coherence.

Evaluating now the spatial degree of the transverse coherence at the detector position $\varsigma_D^{DC}$ that was defined in Eq. (12) in the case of *partially coherent* illumination of the grating, we obtain

$$\varsigma_D^{DC} = \left[ \frac{\hat{\alpha} - \hat{\beta}}{\hat{\alpha} + \hat{\beta}} \right]^{1/2} = \left[ \frac{q_t^2 q_c^2 q_{sl}^2 + [4(q_t^2 + q_c^2) + q_t^2 q_c^2]}{(1 + q_{sl}^2)[4(q_t^2 + q_c^2) + q_t^2 q_c^2]} \right]^{1/2}. \qquad (93)$$

We see from this expression that when the slits are open, $q_{sl} \Rightarrow \infty$, we have for the degree of coherence our previous partial coherent result stated in Eq. (61) and when the slits are rather closed, $q_{sl} \Rightarrow 0$, we obtain for the degree of coherence the just value of one or hundred percent of coherence.

By closing the slits, we obtain a delta-function point source with a certain bandwidth determined by a resolution function. As such, we have a point source with perfect spatial coherence, but slightly spoiled longitudinal coherence. We define the spectral resolution function $R(\omega)$ as

$$R(\omega) = exp\left[ -\frac{(\Delta\omega)^2}{2(\sigma_{res}^{\omega})^2} \right], \qquad (94)$$

where the resolution of the spectrometer $\sigma_{res}^{\omega}$ in the *completely coherent* illumination case is defined in Eq. (70). In this case, we obtain for the CDC $\gamma(\tau)$ (neglecting spatial coordinates)

$$\gamma(\tau) = \frac{\Gamma(\tau)}{\Gamma(0)} = e^{-i\omega_0\tau} exp\left[-\frac{\tau^2}{2l_\tau^2}\right], \qquad (95)$$

where the longitudinal coherence length $l_\tau$ is defined as

$$l_\tau = \frac{1}{\sigma_{res}^\omega}. \qquad (96)$$

## VII. Discussion

We consider a beamline with an undulator source and a plane VLS grating as shown in Fig. 3. This beamline is a simplified version of the beamline P04 at the PETRA III synchrotron storage ring at DESY in Hamburg. The parameters that are regarded in our case are similar to ones considered by Ref. [20]. The distance from the undulator center to the grating is taken to be $L = 45$ m and from the grating to the exit slits $f = 25$ m. We consider the photon energy $E_0 = 1.2$ keV. We assume that the undulator beam has a narrow photon energy bandwidth of $\Delta E = 0.2$ eV that satisfies the condition $\Delta E/E_0 = 1.7 \cdot 10^{-4} \ll 1$, meaning that the quasi-monochromatic approximation is valid for our x-ray beam. The grating period $\kappa = 1/d_0 = 1200 \, l/mm$. We will consider the scattering of the incoming field on a VLS grating in reflection geometry as depicted in Fig. 3. The incidence angle on the grating calculated from the normal to the grating surface is considered to be $\alpha = 88.35°$ and the exit angle in the first order is $\beta_1 = 86.7°$. We also assume that the incoming beam profile is Gaussian with the FWHM equal to 500 μm or with the rms value of $\sigma_{in}^x = 212.3$ μm.

We will first check if the condition in Eq. (8) is well satisfied with these parameters of the grating and the beam. A simple estimate shows that the time delay $\tau$ should satisfy the following condition

$$\tau = \frac{(\Delta x)^2}{2cf} \ll \frac{1}{\Delta\omega}. \qquad (97)$$

That means that the separation between two points at the focal plane of the VLS grating should be

$$\Delta x \ll \left[(2cf)\left(\frac{1}{\Delta\omega}\right)\right]^{1/2}. \qquad (98)$$

Our estimate for the right side of the inequality in Eq. (95) with the parameters given above gives 7 mm. That means that the condition in Eq. (8) is very well satisfied in our case.

We present now results obtained from these parameters and a *completely coherent* beam incoming on the VLS grating (see Eq. (92)). In Fig. 4 the degree of coherence $\varsigma^{DC}$ is shown (red line) as a function of the slits opening. First, we see from this figure that at comparably open slits of 100 μm the degree of coherence is on the order of 8.65%. As soon as we close the slits the degree of coherence is slowly increasing and finally is coming to 100% (we should note here that in the case of rather closed slits in Ref [20] it was a bug in calculations that was corrected in the forthcoming paper [27]). We see that we started with the fully coherent beam but finally obtained the beam that is partially coherent from 8.65% to 100% depending on the size of the exit beam slits. We also checked how the degree of coherence will look like if we change the parameter $q_c$ by making it ten times higher (blue dashed curve) and ten times lower (green dash-dot curve). We notice that for open slits, when the parameter $q_c$ is higher we have about 50% of the degree of coherence (47.12%) and when this parameter is lower we obtain an even lower value of 6.88% for the degree of coherence. It is interesting that all curves give 100% of the degree of coherence with rather closed slits.

## VIII.    Conclusions

In this work, first, the amplitude of the electric field in the focus of the VLS grating was obtained under quasi-monochromatic conditions, $\Delta\omega/\omega_0 \ll 1$. We noticed here that the amplitude of the field provides spatial-frequency coupling after scattering from the grating. After that we calculated correlation functions at the position of the exit slits of the grating assuming *completely coherent* illumination of the grating. We found that the grating was suppressing coherence in its dispersive direction. We determined that the degree of coherence can be substantially lower than 100%, even when the incoming beam is fully coherent. After that we assumed that the beam incoming on the grating is *partially coherent* and obeying a Gaussian Schell-model beam. We determined that in this case the degree of coherence at the exit slits position is additionally degraded. We, finally, determined the degree of coherence for different sizes of the exit slits assuming that $\Delta\omega/\omega_0 = 1.7 \cdot 10^{-4}$. We obtained that in this case the degree of coherence is about 8.65% in the case of open slits and about 100% in the case of rather closed slits, although the incoming beam on the grating was completely coherent.

We would like to note also that the same results may be obtained for the plane grating with the constant line spacing equipped with an additional focusing mirror. In this case the distance *f* will be just the focusing length of this mirror.

All these findings may be significant for the 4th generation diffraction limited synchrotron sources, which are expected to be almost entirely coherent in soft x-ray region up to 1-2 keV [6].

**APPENDIX I**

We start with the scattering on a grating in reflection geometry (see Fig. 1) in the frame of the 1st-order Born approximation following the outline of the known approach [26]. First, we will assume that the grating is not vibrating, that means that we will assume static scattering. Second, we consider a monochromatic x-ray field with time dependence $exp(-i\omega t)$ incident on a grating occupying a finite domain $V$ (in the following we will omit the time dependence). Assuming also a scalar x-ray field in a spatial-frequency domain $E(\boldsymbol{r}, \omega)$ we consider the full wavefield as a sum of the incoming field $E'_{in}(\boldsymbol{r}, \omega)$ and of the scattered field $E_s(\boldsymbol{r}, \omega)$

$$E(\boldsymbol{r}, \omega) = E'_{in}(\boldsymbol{r}, \omega) + E_s(\boldsymbol{r}, \omega) . \qquad (AI, 1)$$

The incident field we take in the following form

$$E'_{in}(\boldsymbol{r}, \omega) = E_{in}(\boldsymbol{r}, \omega) e^{i\boldsymbol{k}\cdot\boldsymbol{r}} , \qquad (AI, 2)$$

where $\boldsymbol{k} = k\boldsymbol{s}_0$, $k = \omega/c$, $c$ is the speed of light, $\boldsymbol{s}_0$ is a unit vector in the direction of the incoming momentum vector and dot means a scalar product between vectors $\boldsymbol{k}$ and $\boldsymbol{r}$. In Eq. (AI, 2) we assume that the factor $exp(i\boldsymbol{k}\cdot\boldsymbol{r})$ is a plane wave and the incoming amplitude $E_{in}(\boldsymbol{r}, \omega)$ is a slowly varying function of its arguments. At the same time, just this incoming amplitude will be fluctuating in time.

For large enough distances $\boldsymbol{r}=r\boldsymbol{s}$, where $\boldsymbol{s}$ is a unit vector, we obtain

$$E_s(r\boldsymbol{s}, \omega) = A(\boldsymbol{s}, \boldsymbol{s}_0; \omega) \frac{e^{ikr}}{r} , \qquad (AI, 3)$$

where the amplitude of scattering is

$$A(\boldsymbol{s}, \boldsymbol{s}_0, \omega) = \left(\frac{k^2}{4\pi}\right) \int \chi(\boldsymbol{r}', \omega) E(\boldsymbol{r}', \omega) e^{-ik\boldsymbol{s}\cdot\boldsymbol{r}'} d\boldsymbol{r}' . \qquad (AI, 4)$$

Here $\chi(\boldsymbol{r}, \omega)$ is the susceptibility of the grating and this equation will be valid if the incoming field $E_{in}(\boldsymbol{r}, \omega)$ satisfies the following condition

$$\boldsymbol{k} \cdot \frac{\partial E_{in}(\boldsymbol{r}, \omega)}{\partial \boldsymbol{r}} = 0 . \qquad (AI, 5)$$

Now, in order to obtain a solution for the amplitude of scattering $A(s, s_0, \omega)$ we will use the so-called 1st-order Born approximation. We will assume, as usual in x-ray science, that scattering by the grating is weak and we will substitute the total field $E(r, \omega)$ in Eq. (AI, 4) by its value in Eq. (AI, 2)

$$A(q, \omega) = \frac{k^2}{4\pi} \int \chi(r', \omega) E_{in}(r', \omega) e^{-iq \cdot r'} dr', \qquad (AI, 6)$$

where $q = k(s - s_0)$ and is valid when $kr \gg 1$. As we are looking on a scattering sphere at far distances, we can substitute the scattering amplitude $A(q, \omega)$ by the scattered field $E_s(q, \omega)$ and, in addition, for scattering from a grating we will assign the corresponding subscript to this scattered field $E_{gr}(q, \omega)$.

We will now assume that the scattering process is two-dimensional and is performed in the *x,z*-plane. We will also assign that in the incoming and outgoing beams the *z*-direction is along the beam propagation and the *x*-direction is perpendicular (see Fig. 1). We will also assume that on the grating the *z*-coordinate is perpendicular to the grating and the *x*-coordinate is along the grating. The scattered amplitude from the grating may be obtained at the exit surface of this grating according to the following recipe [26,28] $\chi(r, \omega) \rightarrow R_{gr}(x)$, where $R_{gr}(x)$ is the reflection function of the grating with the grooves considered along the *y*-direction and we assume also that the grating response does not depend on the frequency $\omega$. After all this, we obtain

$$E_{gr}(q_x, \omega) = E_0 \int R_{gr}(x') E_{in}(x', \omega) e^{-iq_x \cdot x'} dx', \qquad (AI, 7)$$

where the integration in this equation is performed along the grating. In this equation, $E_0$ is the amplitude of the field in which we will further incorporate all nonessential pre-integral factors.

We will introduce now the finite size grating function as in Ref. [28]

$$\begin{aligned} R_{gr}(x) &= \sum_{n=-N/2}^{N/2} r_{gr}(x - x_n) = R_N(x) \sum_{n=-\infty}^{\infty} r_{gr}(x - x_n) \\ &= R_N(x) \sum_{h_n=-\infty}^{\infty} r_{gr}(h_n) e^{ih_n x}, \end{aligned} \qquad (AI, 8)$$

where N is the number of grating periods, $r_{gr}(x)$ is the reflection function of one period, $x_n = d_0 n$ with $n = 0, \pm 1, \pm 2, \cdots$, and $d_0$ is the spacing of the grooves of the grating. In Eq. (AI, 8) $R_N(x)$ is a finite window of the length of the grating, that can be represented as

$$R_N(x) = rect\left(\frac{x}{D_N}\right), \tag{AI, 9}$$

where $D_N = d_0 N$ and $rect(x)$ is a rectangular function defined as

$$rect(x) = \begin{cases} 1, & |x| \leq 1/2 \\ 0, & |x| > 1/2 \end{cases}, \tag{AI, 10}$$

In Eq. (AI, 8) we used also the decomposition of the infinite periodic function over reciprocal space with $h_n = (2\pi/d_0)n$, and introduced the Fourier transform of the reflection function of one grating period $r_{gr}(x)$ as

$$r_{gr}(h_n) = \frac{1}{d_0} \int_{-d_0/2}^{d_0/2} r_{gr}(x) e^{-ih_n x} dx. \tag{AI, 11}$$

Substituting now Eq. (AI, 8) in Eq. (AI, 7) and changing the order of summation and integration we obtain

$$E_{gr}(q_x, \omega) = E_0 \sum_{h_n=-\infty}^{\infty} r_{gr}(h_n) \int R_N(x') E_{in}(x', \omega) e^{-i(q_x-h_n)\cdot x'} dx'. \tag{AI, 12}$$

We have to note here that the parameter $q_x$ is defined by the scattering conditions (see Fig. 1) [26]

$$q_x = k(s_x - s_{0x}) = k(\sin\beta_n + \sin\alpha). \tag{AI, 13}$$

Now, if we compare Eq. (AI, 12) with the corresponding expression in Ref. [28], we see that there is no difference except the fact that we assumed while deriving Eq. (AI, 12) that the distances satisfy the condition $kr \gg 1$ and in Ref. [28] the exit plane was taken in the focal plane $f$ of the VLS grating. Taking this into account, we will use Eq. (AI, 12) also for the focal plane of the VLS grating.

In the following we will assume that the grating size $D_N$ is much larger than the illuminating field and we will consider scattering to the *n*-th order. Taking all this into account, we get from Eq. (AI, 12)

$$E_{gr}(q_x, \omega) = E_0 r_{gr}(h_n) \int E_{in}(x', \omega) e^{-i(q_x-h_n)\cdot x'} dx'. \tag{AI, 14}$$

**APPENDIX II**

In the text of the paper the following known integral was used many times

$$\int_{-\infty}^{\infty} e^{-\beta^2 t^2} e^{-iqt} dt = \frac{\sqrt{\pi}}{\beta} exp\left[-\frac{q^2}{4\beta^2}\right]. \qquad (AII, 1)$$

By using the same integral, it is possible also to derive (see, for example, Ref. [22]) that

$$J(q_{x_1}, q_{x_2}) = J_0 \iint dx' dx'' e^{-(ax'^2 + ax''^2 - 2bx'x'')} e^{-i(q_{x_2}x'' - q_{x_1}x')}$$
$$= J_0 e^{-(\alpha q_{x_1}^2 + \alpha q_{x_2}^2 - 2\beta q_{x_1} q_{x_2})}, \qquad (AII, 2)$$

where

$$\alpha = \frac{a}{4(a^2 - b^2)} \ ; \ \beta = \frac{b}{4(a^2 - b^2)}. \qquad (AII, 3)$$

Considering that according to Eq. (31) $q_{x_{1,2}} = (k_0 \cos \beta_n / f)(x_{1,2} - x_n)$ we can rewrite Eq. (AII, 2) in the following form

$$J(\Delta x_1, \Delta x_2) = J_0 exp\{-[\tilde{\alpha}\Delta x_1^2 + \tilde{\alpha}\Delta x_2^2 - 2\tilde{\beta}\Delta x_1 \Delta x_2]\}, \qquad (AII, 4)$$

where

$$\tilde{\alpha} = \alpha_0^2 \alpha \ ; \ \tilde{\beta} = \beta_0^2 \beta \ ; \alpha_0^2 = \beta_0^2 = k_0^2 \cos^2 \beta_n / f^2 \ \ \Delta x_{1,2} = x_{1,2} - x_n \qquad (AII, 5)$$

we can now determine all parameters that are of interest for us.

For example, we can determine that from Eq. (11) we have for the intensity

$$I(\Delta x) = I_0 exp\{-2(\tilde{\alpha} - \tilde{\beta})\Delta x^2\} = I_0 exp\left[-\frac{\Delta x^2}{2(\sigma_I)^2}\right], \qquad (AII, 6)$$

where

$$\sigma_I = \frac{1}{2\sqrt{\tilde{\alpha} - \tilde{\beta}}}. \qquad (AII, 7)$$

For the CDC $j(\Delta x_1, \Delta x_2)$ we obtain according to its definition in Eq. (9) and Eq. (AII, 4)

$$j(\Delta x_1, \Delta x_2) = exp[-\tilde{\beta}(x_2 - x_1)^2] = exp\left[-\frac{(x_2 - x_1)^2}{2L_c^2}\right], \qquad (AII, 8)$$

where

$$L_c = \frac{1}{\sqrt{2\tilde{\beta}}}. \qquad (AII, 9)$$

Finally, for the spatial degree of the transverse coherence that was defined in Eq. (12), by performing the necessary integration of Eqs. (AII, 4) and (AII, 6) we obtain

$$\varsigma^{DC} = \sqrt{\frac{\tilde{\alpha} - \tilde{\beta}}{\tilde{\alpha} + \tilde{\beta}}}. \qquad (AII, 10)$$

**Acknowledgements**

The author has appreciated numerous discussions with Kai Bagschik and Ruslan Khubbutdinov who pointed out to the problem of reduced spatial coherence after scattering on the VLS grating. The author also appreciated a number of discussions with C. Schroer. The author is thankful to M. Hoesch and G. Hinsley for careful reading of the manuscript.

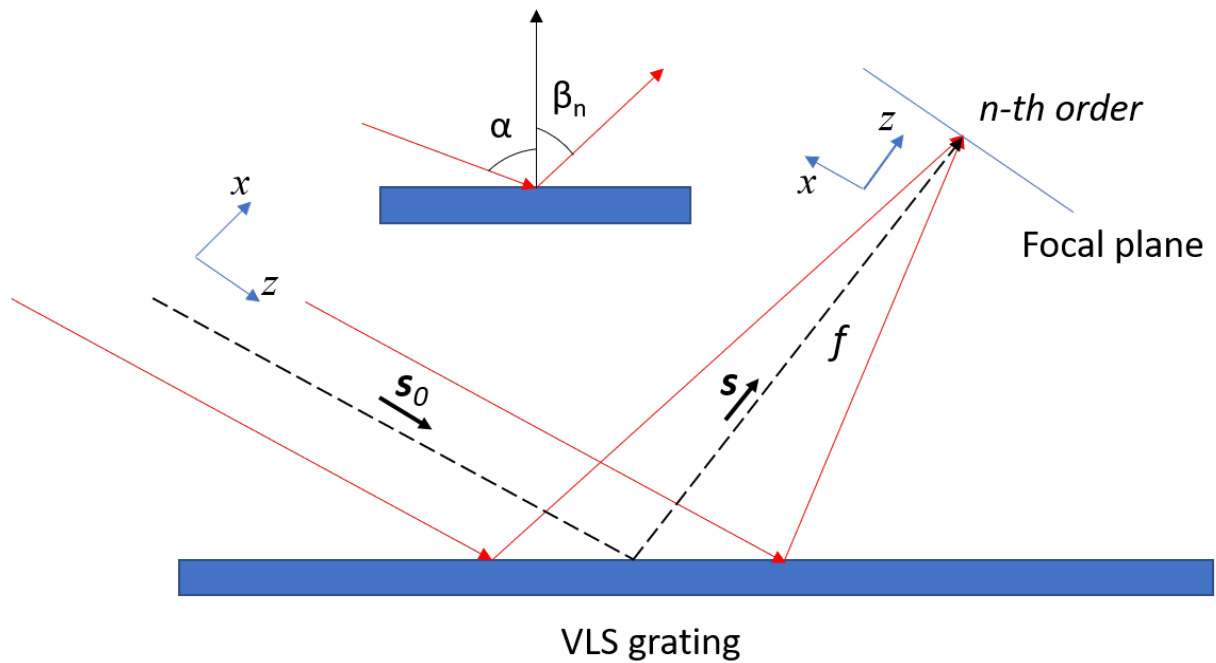

**Fig. 1.** Scattering of soft x-ray radiation on a VLS grating. The incoming x-ray radiation with the direction given by the unit vector $\mathbf{s_0}$ is focused by a VLS grating to a focal plane at a distance $f$. The scattered radiation is going to the $n$-th order of the grating defined by the unit vector $\mathbf{s}$. The coordinate system is considered to be two-dimensional and is performed in the $x,z$-plane with the $z$-direction being along the beam propagation direction. The incoming $\alpha$ and outgoing $\beta_n$ angles are taken from the normal of the grating as shown in the inset.

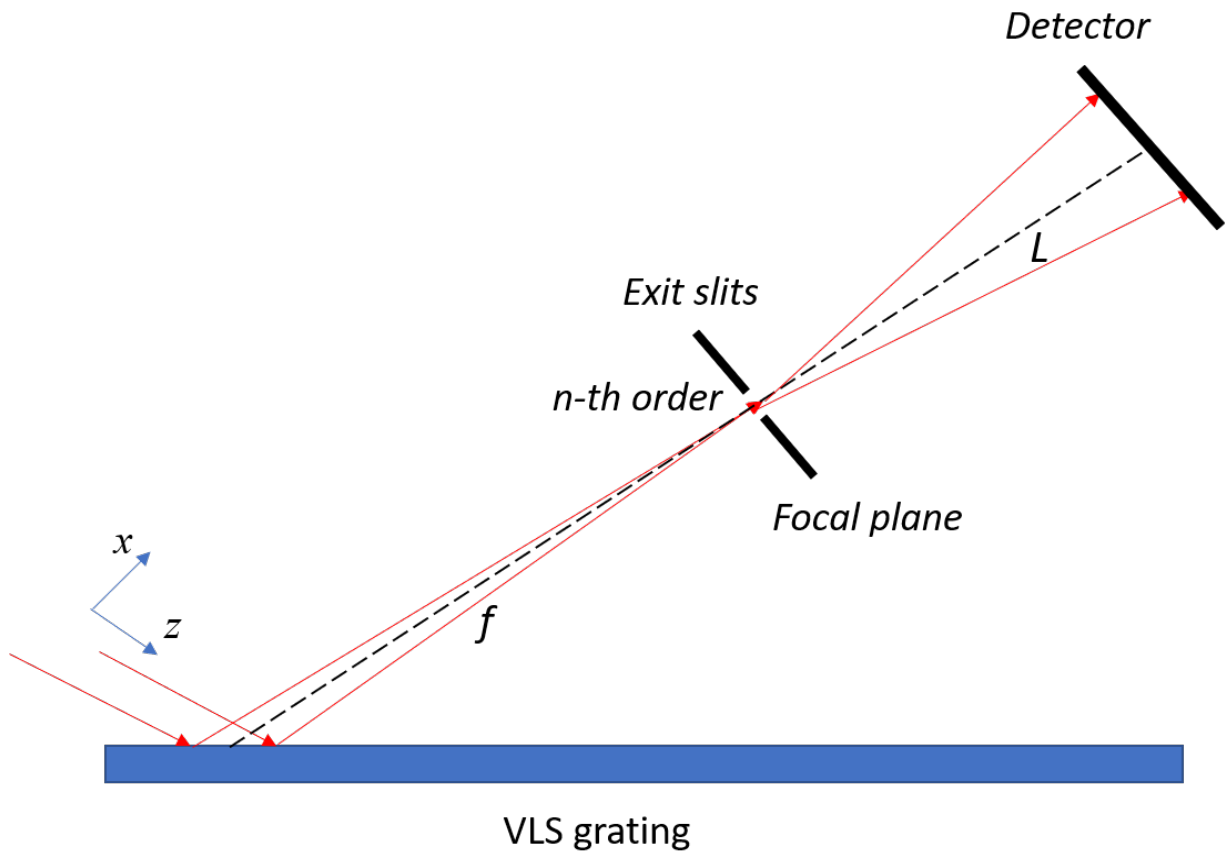

**Fig. 2.** Monochromator configuration of the VLS grating. In this configuration, exit slits of varying sizes are positioned at the focal plane of the grating at a distance $f$. These slits define the bandwidth of radiation that is passing through. After the slits, at a distance $L$, we assume a detector at which coherence measurements are taking place.

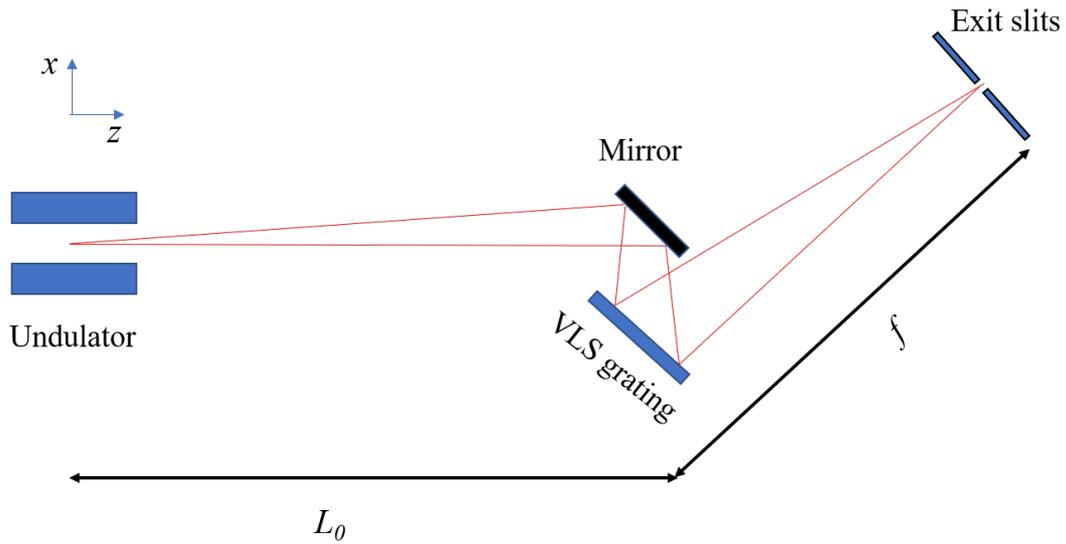

**Fig. 3**. Representation of a simplified set-up of the P04 beamline at PETRA III. X-ray radiation is produced by the undulator source then scattered on the mirror which illuminates a VLS grating. After scattering from the grating, the x-ray radiation is focused to the position of the exit slits. The distance from the undulator source to the grating is assumed to be $L_0$=45 m and the distance from the grating to the focal plane is $f$=25 m.

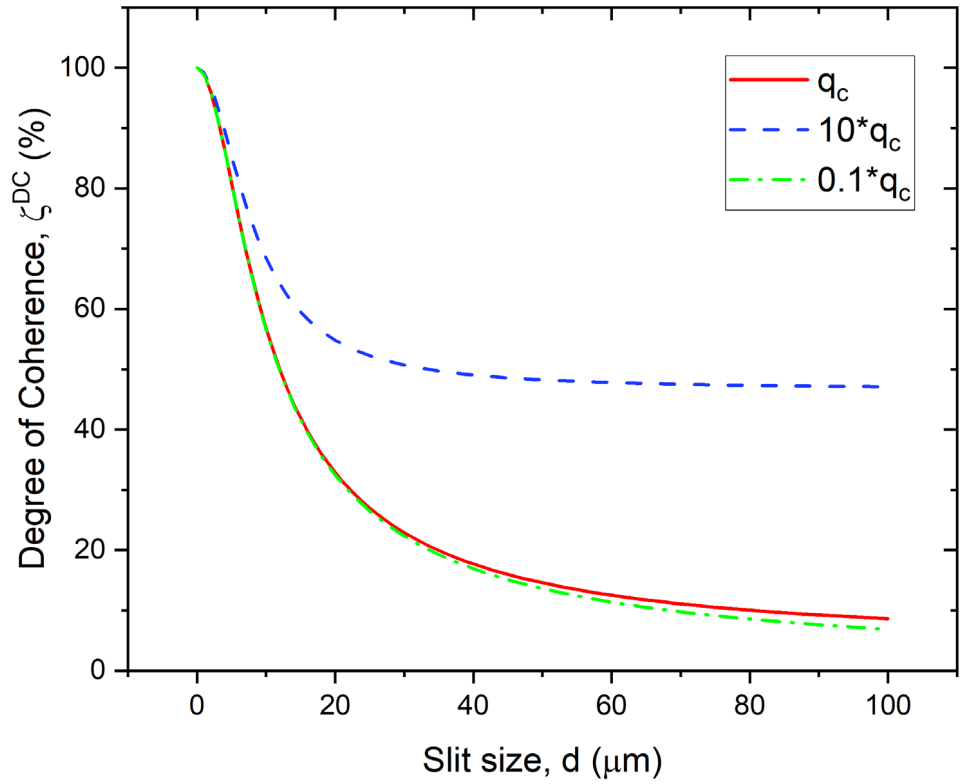

**Fig. 4** Degree of coherence as a function of the slit openings. Here, the red solid line corresponds to the parameters of the simulations that are listed in the paper, the blue dashed line corresponds to the situation when the parameter $q_c$ is ten times higher and the green dash-dot line corresponds to the situation when the parameter $q_c$ is ten times smaller.